\documentclass{PoS}
\usepackage{ifpdf}
\usepackage{calc}
\include{dvipsnam.def}
\usepackage{epsfig}
\usepackage[latin1]{inputenc}
\usepackage{textcomp}
\usepackage{multicol}
\usepackage{amsmath}
\usepackage{amssymb}

\newcounter{figurenum}
\PoS{PoS(LAT2005)150}
\title{Thermodynamics using p4-improved staggered fermion action on QCDOC}
\ShortTitle{Thermodynamics using p4-improved staggered fermion action on QCDOC}

\author{\speaker{Chulwoo Jung} for the RBC-Bielefeld Collaboration\\
Brookhaven National Laboratory and Columbia University, USA\\
E-mail: \email{chulwoo@bnl.gov}
} 
\abstract{We present an exploratory study of the thermodynamics of $N_f=3$ 
QCD with an improved staggered fermions using the QCDOC supercomputer. We 
use a p4 action with MILC-style smeared links (Fat 7). Some details of the 
implementation of the p4 action on QCDOC are discussed and performance 
benchmarks are given. We show preliminary results for the quark mass 
dependence of the pseudo-critical temperature $T_c$ from several lattice volumes. We also make a comparison between p4fat7 and the old p4 action.}
\FullConference{XXIIIrd International Symposium on
Lattice Field Theory\\
 25-30 July 2005\\
Trinity College, Dublin, Ireland}
\begin{document}
\section{Introduction}
Lattice QCD is an important tool for the study of finite 
temperature QCD. Typically one is forced to use larger lattice spacing $(a)$ 
for finite temperature quantities than zero temperature quantities.
This is mainly because the temperature regime of interest 
$T_c = (N_\tau a)^{-1}$ ($N_\tau =$ 4 - 8 ) is about 
(1 fm)$^{-1}$, and simulations at several gauge couplings have to be performed
to locate the transition temperature. 
%This is because of 
%the nature of moleculardynamics algorithm where the autocorrelation time increases near the phase transition, and the need to simulate for a multitude of 
%gauge couplings.

%This makes improved variants of staggered fermion practical, which also 
%preserves U(1) subgroup of chiral symmetry even at large lattice spacing. 
This makes improved staggered fermion formulations attractive
for thermodynamics studies. Improvements of rotational symmetry via "knight's 
term" (p4 action) largely reduces the cut-off effects in the pressure and 
energy density at high temperatures \cite{Heller:1999xz}.
Also, it is well known that staggered fermions break the flavor symmetry
of continuum QCD and this effect can be largely reduced if fat links
are used in the fermion action \cite{Orginos:1999cr}.
%Here we study a combination of  p4 action \cite{Karsch:2000ps} , which has improved rotational symmetry over standard staggered
% fermion action, with fat7 smeared 1-link term, similar to the one used in 
%asqtad action\cite{Orginos:1999cr} which improves flavor symmetry breaking (p4fat7 action).
% This action is hoped to have a better flavor symmetry than the action used in \cite{Engels:1996ag,Karsch:2000kv}, which had only 3 staple term, which we denote as p4fat3. 
%Some comparison between p4fat7 and the p4 action used in \ref{} is described in section \ref{section:p4fat3}.
%has been used for QCD thermodynamics by Bielefeld group. 
In this contribution we present an
exploratory study of QCD phase diagram with a fermion action which combines 
the improvements in \cite{Heller:1999xz} and \cite{Orginos:1999cr}(p4fat7 
action).
A more detailed description of the action and simulation parameters are given in section \ref{section:action}.
The phase transition for 3 degenerate flavors of p4fat7 quark is described in section \ref{section:Nf3}. Comparison between p4fat7 and p4 action used in 
\cite{Karsch:2000ps} is described in section \ref{section:p4fat3}.
%The relative coefficient between fat7 smearing and p4 is fixed to ???.
%\vspace{-1cm}

\section{Action and simulation parameters}
\label{section:action}
For this study we use the tree-level improved Symanzik gauge action and p4fat7 action.
\begin{gather}
S(x) = \beta S_g(x) + S_f(x), \nonumber \\
S_g(x) =\sum_{\mu > \nu} \left[\frac53\left(1 -\frac13 \mbox{ReTr} [U_\mu(x) U_\nu(x+\hat{\mu}) U^\dagger_\mu(x+\hat{\nu}) U^\dagger_\nu(x)]
 \right) \right. \\
\left. -\frac16\left(1 - \frac16 \mbox{ReTr} \left[U_\mu(x) U_\mu(x+\hat{\mu}) U_\nu(x+2\hat{\mu}) U^\dagger_\mu(x+\hat{\mu} + \hat{\nu}) U^\dagger_\mu(x+\hat{\nu}) U^\dagger_\nu(x)\right]-\frac16 \mbox{ReTr}[ \mu \leftrightarrow \nu]\right) \right]. \nonumber
\end{gather}
\begin{figure}[h]
\begin{center}
\begin{minipage}{0.6\linewidth}
\includegraphics[width=\textwidth,bb= 0 0 563 220,clip]{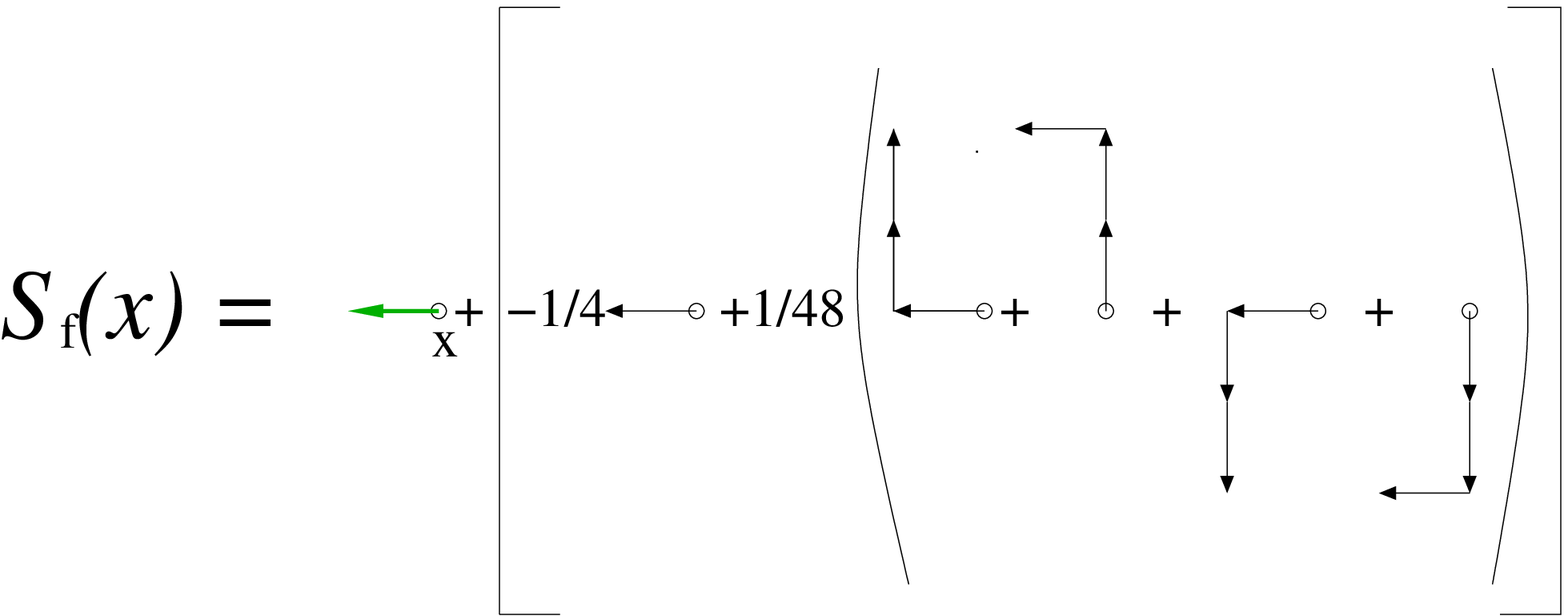}
\includegraphics[width=\textwidth,bb=0 0 590 260,clip]{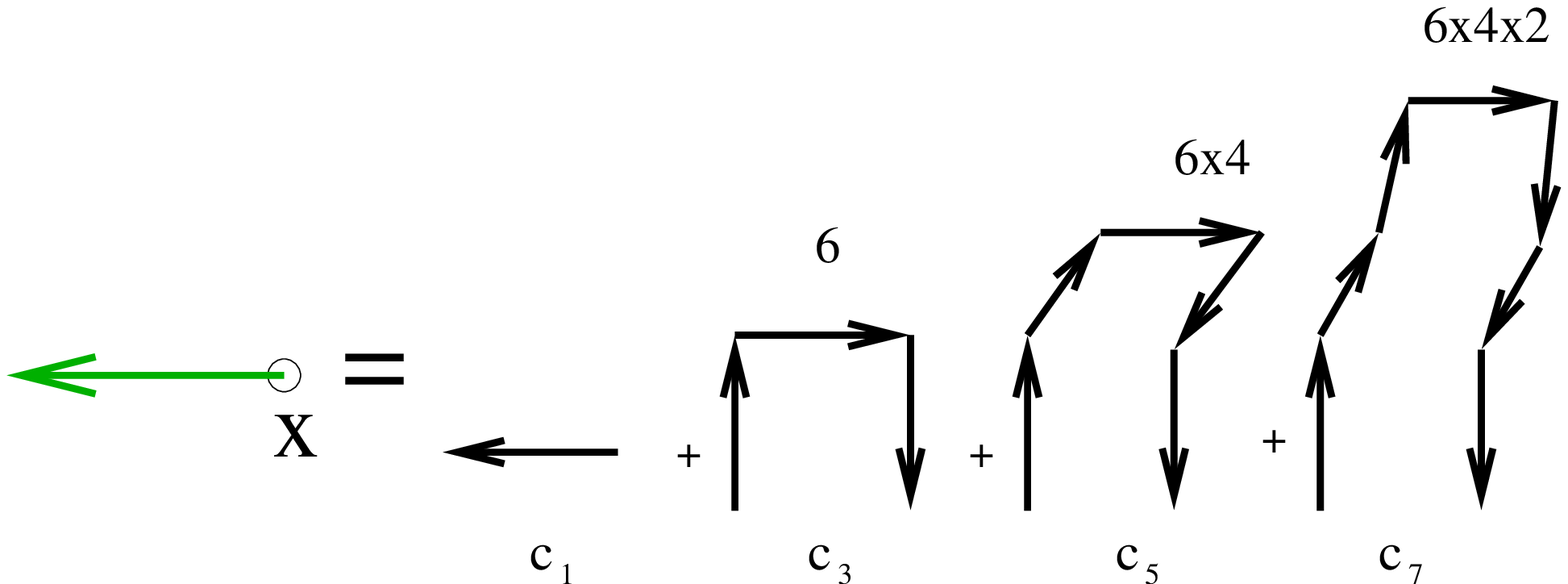}
\end{minipage}
\end{center}
%\begin{minipage}{0.27\linewidth}
\caption{Diagram of p4fat7 action. The numbers above graphs denote the multiplicities of each graphs. } 
%\end{minipage}
\label{figure:p4fat7}
\end{figure}
%In Figure \ref{figure:p4fat7}, the tree-level coefficients are
The fermion part of the action ($S_f$) is shown in Fig. \ref{figure:p4fat7}.
The coefficient in the smeared link are chosen to
cancel flavor symmetry breaking at order $\alpha_s a^2$~\cite{Orginos:1999cr},
\begin{equation}
c_1 = \frac{1}{8},\quad\
 c_3 = \frac{1}{16},\quad\
 c_5 = \frac{1}{64},\quad\
 c_7 = \frac{1}{384}.
\end{equation}
Here we use the above values of the fat7 coefficients without tadpole improvement.
% with tree-level fat7 smearing coefficients instead of tadpole-improved ones.
%We do not use tadpole improvement.
%For a more detailed discussion, please refer to \cite{Cheng}.
The effect of tadpole improvement on quenched configurations is studied in \cite{Cheng}. 
 
\begin{figure}[h]
\begin{center}
%\begin{minipage}{0.7\linewidth}
%\includegraphics[width=3cm,bb= 0 0 420 192,clip]{p4_2.eps} 
\epsfig{file=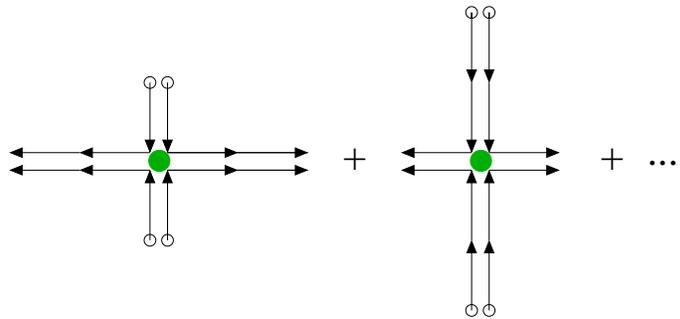,width=0.6\textwidth,bb= 0 0 420 192,clip}
%\end{minipage}
\end{center}
%\begin{minipage}{0.25\linewidth}
\caption{Illustration of implementation of p4 action. The arrows and circles denote the parallel transports and recombinations respectively.} 
%\end{minipage}
\label{figure:recom}
\end{figure}

All the codes needed for dynamical gauge evolution (HMD R algorithm) is implemented  for QCDOC in Columbia Physics System (CPS) and numerically checked against existing Asqtad and p4 action results.
The "knight's move" term of p4 action is implemented without non-nearest communication by breaking the term into
Parallel transport (  $ {\color{black}\psi'_\mu(x)} = {\color{black} U_\mu(x)} {\color{black} \psi(x+\hat\mu)}$)  
and recombination $\left( {\color{black} R_\mu(x)} = \sum_{\nu \neq \mu} ({\color{black} \psi(x+\hat{\nu}) \pm \psi(x-\hat{\nu})}\right)$, as illustrated in Fig. \ref{figure:recom}. 

\begin{table}
\begin{center}
\begin{tabular} {p{2cm}|p{2cm}|p{4cm}|p{4cm}}
\hline
Volume&	sites/node&CG(MFlops/node) & Smearing(MFlops/node)\\
\hline
%2224&	32&	124&	241\\
$2^24^2$&	64&	173&	271\\
%$2\times 4^3$&	128&	177&	293\\
$4^4$&	256&	207&	313\\
%$4^3\times 6$&	384&	254&	317\\
$4^2\times 6^2$&	576&	264&	264\\
%$4\times6^3$&	864&	267&	181\\
$6^4$&	1296&	231&	164\\
%$6^3\times 8$&	1728&	231&	166\\
\hline
\end{tabular}
\end{center}
\label{table:perf}
\caption{Performance of p4fat7 evolution codes on QCDOC,
420Mhz, 1024 nodes, MFlops/node}
\end{table}

%Table \ref{table:perf} shows the performance of Dirac operator and the fat7 smearing routine for p4fat7. 
Inversion of p4fat7 Dirac operator, which is the dominating part of gauge evolution for small masses, is currently running at $\sim $ 31\% of 
the peak for $4^3\times 6$ local volume, which is the local volume for 
$32^3\times 6$ on a 512-node QCDOC partition. Further optimization of the Dirac operator, especially at smaller volumes, is possible.
Rational Hybrid Monte Carlo(RHMC)\cite{Clark:2003na} implementation is also in progress.
All the simulations were done with R algorithm\cite{Gottlieb:1987mq} with the step size equal to 40\% of the bare mass in lattice units.
\boldmath
\section{$N_f=3$ phase transition with p4fat7}
\unboldmath
\label{section:Nf3}
\begin{figure}[h]
\begin{minipage}{0.5\linewidth}
\includegraphics[width=\textwidth,bb=50 50 410 302]{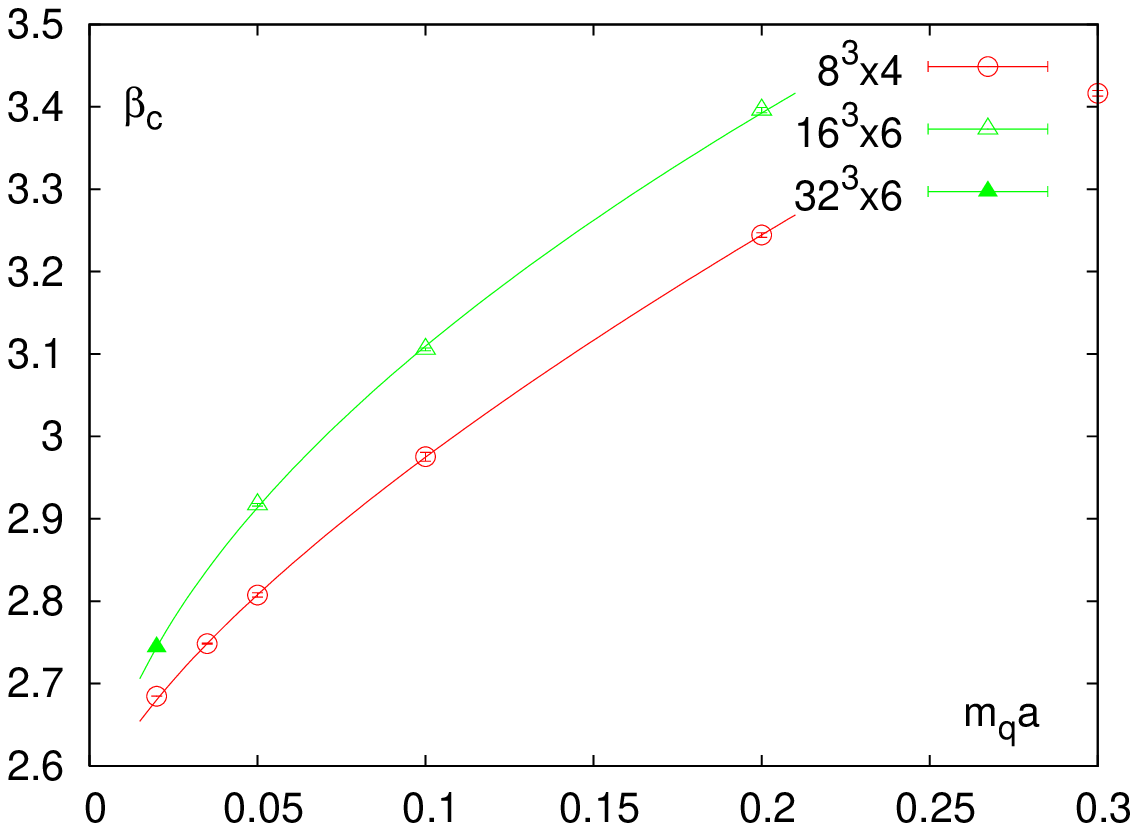}
\end{minipage}
\begin{minipage}{0.5\linewidth}
\includegraphics[width=1.05\textwidth,bb= 50 50 410 302]{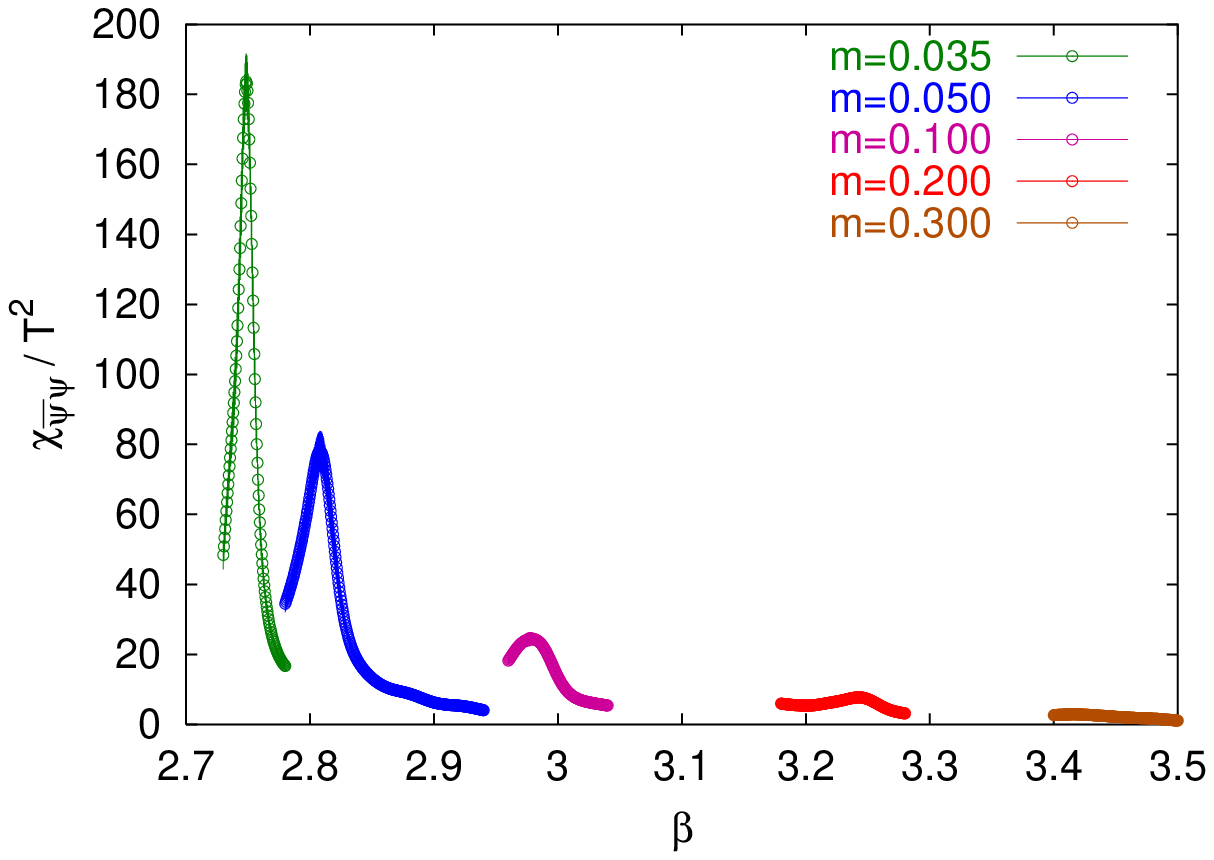}
\end{minipage}
\begin{minipage}{0.45\linewidth}
\label{gr:betac}
\caption{Transition coupling in lattice units ($\beta_c$) for 3 flavor p4fat7 
action. 
The curves represent fits to $\beta_c(m_q) = A+ B (m_q a)^C$.
Red and green points denote $N_\tau=4$ and $N_\tau=6$ respectively. }
\end{minipage}
\hspace{0.1\linewidth}
\begin{minipage}{0.45\linewidth}
\caption{The susceptibility in chiral condensate for $8^3\times 4$ }
\label{gr:suscept}
\end{minipage}
\end{figure}
Fig. \ref{gr:betac} shows the coupling at the transition as a function of quark masses in lattice units. 
$\beta_c$ are located by measuring susceptibilities in average plaquette, Polyakov loop and chiral condensate. The location of peaks coincide within errors.
The use of tree-level coefficients enables the use of the
reweighting procedure, which is used in locating the transition temperature.
At $\beta_c$ for each mass, separate measurements of heavy quark potential 
and spectrum calculations were done on $16^3\times 32$ lattices for physical 
scale setting. 

It has been shown that staggered type fermions exhibits a transition from a rapid crossover to a 1st order transition in small masses. 
Initial studies suggest that the mass where the transition occurs for $N_\tau=4$ may be larger than \cite{Karsch:2000ps}, as shown by the rapidly increasing peak in susceptibility in Fig.~\ref{gr:suscept}.  This may indicate the additional smearing does not suppress finite lattice spacing error much.
 Further study is in progress.

%\end{minipage}

%\vspace{1cm}
%\begin{minipage}[l]{0.15\textwidth}

%\includegraphics[width=10in, bb= 50 50 410 302]{Tc_sigma4.pdf}
%\includegraphics[width=10in, bb= 50 50 410 302]{Tc_r04.pdf}
%\includegraphics[width=10in,bb= 50 50 410 302,clip]{tc_sqrtsigma_pirho_nf3_av.pdf}
\begin{figure}[h]
\begin{tabular}{ll}
\includegraphics[width=0.5\textwidth,bb= 50 50 410 302,clip]{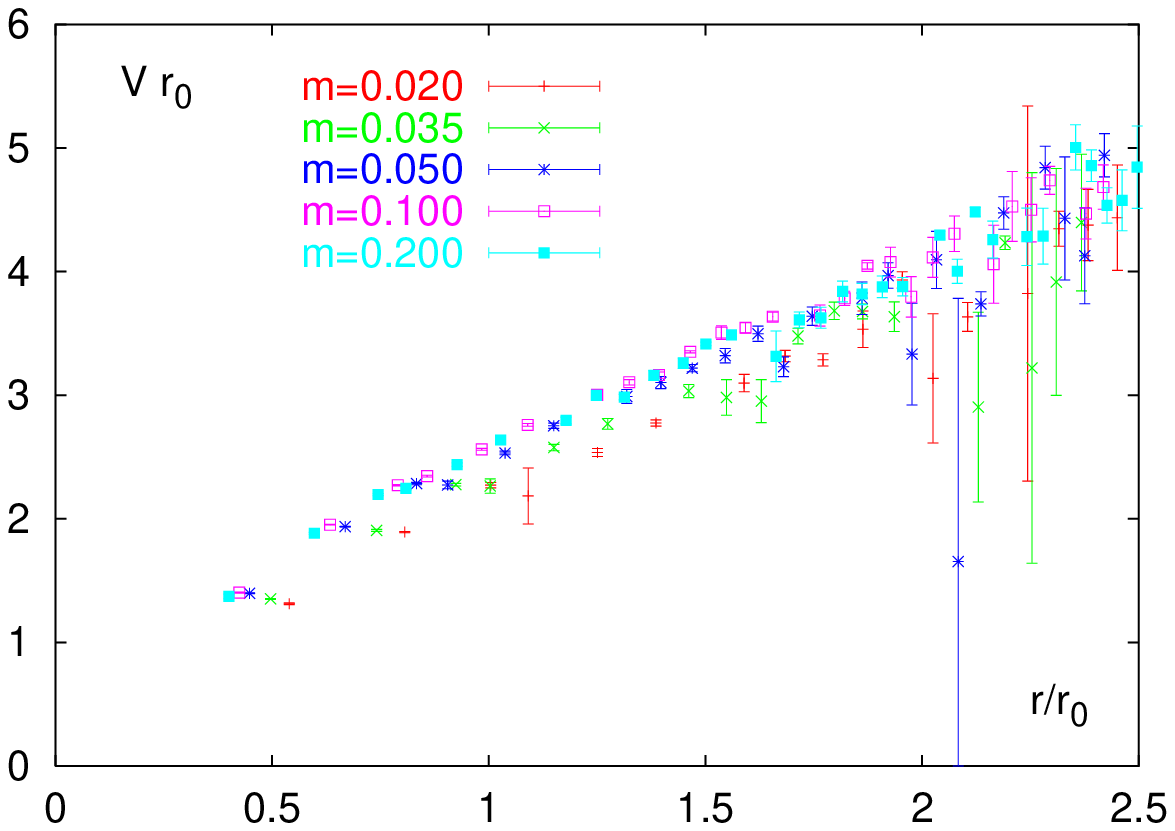}
\includegraphics[width=0.5\textwidth,bb= 50 50 410 302,clip]{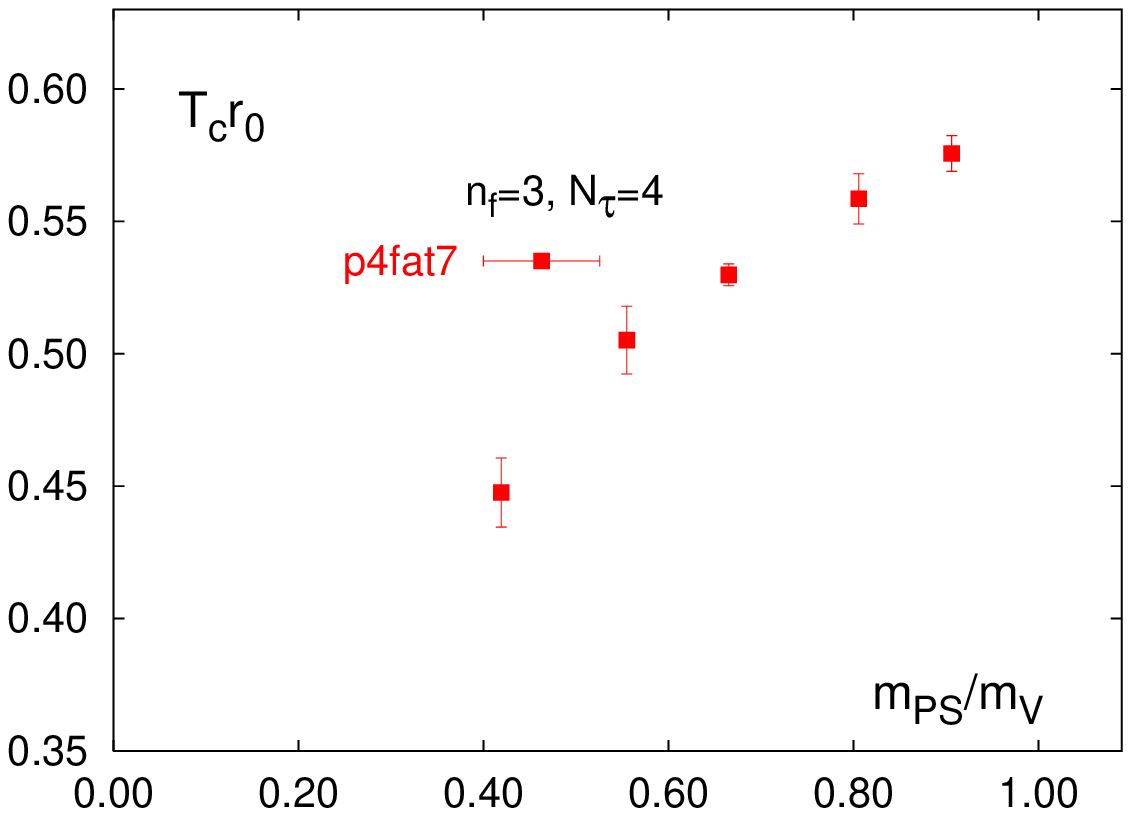}
\end{tabular}
\caption{Scale from heavy quark potential for $8^3 \times 4$ p4fat7 lattices}
\label{gr:pot}
\end{figure}
%\begin{figure}[h]:1
%\includegraphics[width=0.5\textwidth,bb= 50 50 410 302,clip]{tc_sqrtsigma_pirho_nf3_av.eps}
%\includegraphics[width=0.5\textwidth,bb= 50 50 410 302,clip]{tc_r0_pirho_nf3_av.eps}
%\end{tabular}
%\end{figure}
Fig.~\ref{gr:pot} shows the heavy quark potential and the lattice spacing, measured by $r_0$\cite{Guagnelli:1998ud}. The error in each $T_c r_0$ includes the difference between 2 different spacial smearings. 

\section{Comparison with p4fat3 action}
\label{section:p4fat3}
P4fat3 action, used in \cite{Karsch:2000ps}, has the knight's move term, 1-link term as well as 3-staple terms. In the notation in Fig. \ref{figure:p4fat7}, the coefficients are 
\begin{equation}
c_1 = \frac34 \frac{1}{1+6\omega},\quad\ c_3 = \frac34\frac{\omega}{1+6\omega},\quad\ c_5 = c_7 = 0,\quad\ \omega = 0.2.
\end{equation}
\begin{figure}[h]
%\begin{minipage}[l]{0.75\textwidth}
\begin{center}
\includegraphics[width=0.75\textwidth,bb= 50 50 410 302,clip]{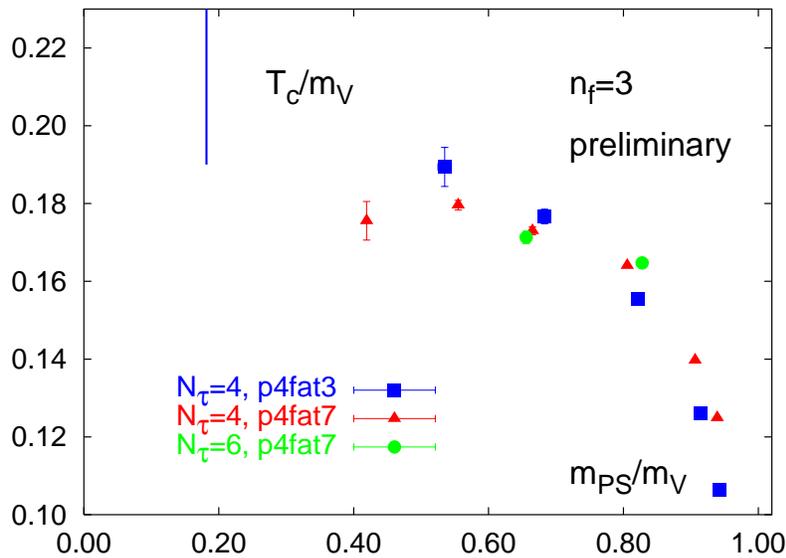}
%\end{minipage}
%\includegraphics[width=0.5\textwidth,bb= 50 50 410 302,clip]{tc_sqrtsigma_pirho_nf3_av.eps}
\end{center}
\caption{Comparison of $T_c/m_V$ between p4fat7 and the p4fat3.}
\label{gr:p4fat3}
\end{figure}

Fig.~\ref{gr:p4fat3} shows the behavior of the transition temperature for the p4fat7 and p4fat3 actions. The transition temperature in units of vector meson mass shows a good agreement between the 2 actions. A further study will be needed for the continuum extrapolation. 

\section{Summary and Future plans}
We studied the phase transition of p4fat7 action for 3 degenerate flavors of 
quarks. p4fat7 action exhibits the finite temperature phase transition and appears to be suitable for other studies such as Equation of States.
Although the use of fat7 smearing shows the expected reduction of flavor
symmetry breaking symmetry breaking in 
smaller lattice spacings\cite{Cheng}, we do not see a large 
improvement in flavor symmetry breaking for the lattice spacing relevant 
for $N_\tau=4$. More improvement such as employing tadpole improvement may be 
needed. 
We also showed some preliminary results for $N_\tau=6$ phase transition. More systematic study of $N_f=3$ and 2+1 transition is in progress.
%The difference between the old p4 action \cite{Karsch:2000ps} and p4fat7 action appears to be at a few percent level. 
Dirac operator and other routines needed for p4fat7 Hybrid Moleculardynamics are optimized for QCDOC and performing at $\sim 30\%$ for the volumes relevant for finite temperature studies.
Implementation for RHMC is under way.

\bibliography{lat05}

\section*{Acknowledgments}
We thank 
Frithjof Karsch, Peter Petreczky, Konstantin Petrov, Christian Schmidt,
Norman Christ, Robert Mawhinney and Michael Cheng for the software 
developments and data analysis for this proceeding, as well as useful and 
enlightening discussions. 

We also thank Peter Boyle, Dong Chen, Norman Christ, Mike Clark, Saul Cohen, 
Calin Cristian, Zhihua Dong, Alan Gara, Andrew Jackson, Balint Joo, 
Chulwoo Jung, Richard Kenway, Changhoan Kim, Ludmila Levkova, Huey-Wen Lin, 
Xiaodong Liao, Guofeng Liu, Robert Mawhinney, Shigemi Ohta, Tilo Wettig, 
and Azusa Yamaguchi for the development of the QCDOC machine and its software. 
This development and the resulting computer equipment were funded
by the U.S. DOE grant DE-FG02-92ER40699, PPARC JIF grant PPA/J/S/1998/00756
and by RIKEN. This work was supported by DOE grant DE-FG02-92ER40699 and
we thank RIKEN, BNL and the U.S. DOE for providing the facilities essential
for the completion of this work.
C.J. is supported by SciDAC project of the U.S. Department of Energy, 
under the contract DE-AC02-98CH10886.

\end{document}